# A Gated Recurrent Unit Approach to Bitcoin Price Prediction

**Aniruddha Dutta** 1, *****, **Saket Kumar** 1 **and Meheli Basu** 2

1 Haas School of Business, University of California Berkeley, CA 94720, USA
2 Joseph M. Katz Graduate School of Business, University of Pittsburgh, PA 15260, USA
***** Correspondence: aniruddha_dutta@berkeley.edu

**Abstract:** In today's era of big data, deep learning and artificial intelligence have formed the backbone for cryptocurrency portfolio optimization. Researchers have investigated various state of the art machine learning models to predict Bitcoin price and volatility. Machine learning models like recurrent neural network (RNN) and long short-term memory (LSTM) have been shown to perform better than traditional time series models in cryptocurrency price prediction. However, very few studies have applied sequence models with robust feature engineering to predict future pricing. In this study, we investigate a framework with a set of advanced machine learning forecasting methods with a fixed set of exogenous and endogenous factors to predict daily Bitcoin prices. We study and compare different approaches using the root mean squared error (RMSE). Experimental results show that gated recurring unit (GRU) model with recurrent dropout performs better than popular existing models. We also show that simple trading strategies, when implemented with our proposed GRU model and with proper learning, can lead to financial gain.

## 1. Introduction

Bitcoin was first launched in 2008 to serve as a transaction medium between participants without the need for any intermediary (Nakamoto 2008; Barrdear 2016). Since 2017, cryptocurrencies have been gaining immense popularity, thanks to the rapid growth of their market capitalization (ElBahrawy 2017), resulting in a revenue of more than $700 billion in 2018. The digital currency market is diverse and provides investors with a wide variety of different products. A recent survey (Hileman, 2017) revealed that more than 1,500 cryptocurrencies are actively traded by individual and institutional investors worldwide across different exchanges. Over 170 hedge funds, specialized in cryptocurrencies, have emerged since 2017 and in response to institutional demand for trading and hedging, Bitcoin futures have been rapidly launched (Corbet 2018).

The growth of virtual currencies (Baronchelli 2018) has fueled interest from the scientific community (Barrdear 2016, Dwyer 2015, Bohme 2015, Casey 2015, Cusumano 2014, Krafft 2018, Rogojanu 2014, White 2015). Cryptocurrencies have faced periodic rise and sudden dips in specific time periods, and therefore the cryptocurrency trading community has a need for a standardized method to accurately predict the fluctuating price trends. Cryptocurrency price fluctuations and forecasts studied in the past (Poyser 2017) focused on the analysis and forecasting of price fluctuations, using mostly traditional approaches for financial markets analysis and prediction (Ciaian 2016, Guo 2018, Gajardo 2018, Gandal 2016). Sovbetov 2018, observes that crypto market-related factors such as market beta, trading volume, and volatility are significant predictors of both short-term and long-term prices of cryptocurrencies. Constructing robust predictive models to accurately forecast cryptocurrency prices is an important business challenge for potential investors and government agencies. Cryptocurrency trading is actually a time series forecasting problem and due to high volatility, it is different from price forecasting in traditional financial markets (Muzammal 2019). Briere et. al

2015 find that Bitcoin shows extremely high return but is characterized by high volatility and low correlation to traditional assets. The high volatility of Bitcoin is well-documented (Blundell-Wignall 2014, Lo 2014). Some econometric methods have been applied to predict Bitcoin volatility estimates such as (Katsiampa 2017, Kim 2016, Kristoufek 2015).

Traditional time series prediction methods include univariate Autoregressive (AR), Univariate Moving Average (MA), Simple Exponential Smoothing (SES), and Autoregressive Integrated Moving Average (ARIMA) (Siami-Namini 2018). Due to the lack of seasonality in the cryptocurrencies market and its high volatility, these methods are not very effective for this task. Machine learning methods seem to be promising in this regard. Machine learning methods have been applied for asset price/return prediction in recent years by incorporating non-linearity in prediction models to deal with non-stationary financial time-series. Machine learning techniques have been successfully applied for stock markets prediction (Enke 2005, Huang 2005, Sheta 2015, Chang 2009). However, there is a dearth of machine learning application in the cryptocurrency price prediction literature. In contrast to traditional linear statistical models such as ARMA, the artificial intelligence approach enables us to capture the non-linear property of the high volatile crypto-currency prices.

Examples of machine learning studies to predict Bitcoin prices include random forests (Madan 2015), Bayesian neural network (Jang 2017), and neural networks (McNally 2018). Deep learning techniques developed by Hinton et. al 2006 have been used in literature to approximate non-linear functions with high accuracy (Cybenko 1989). There are a number of previous works that have applied artificial neural networks to financial investment problems (Chong 2017, Huck 2010). However, Pichl and Kaizoji 2017 conclude that although neural networks are successful in approximating bitcoin log return distribution, more complex deep learning methods such as RNNs and LSTM techniques should yield substantially higher prediction accuracy. Some studies have used RNN's and LSTM to forecast Bitcoin pricing in comparison with traditional ARIMA models (McNally 2018, Guo 2018). McNally 2018 show that RNN and LSTM neural networks predict prices better than the traditional multilayer perceptron (MLP) due to the temporal nature of the more advanced algorithms. Karakoyun 2018 in comparing the ARIMA time series model to the LSTM deep learning algorithm in estimating the future price of Bitcoin, find significantly lower mean absolute error in LSTM prediction.

In this paper, we focus on two aspects to predict Bitcoin price. We consider a set of exogenous and endogenous variables to predict Bitcoin price. Some of these variables have not been investigated in previous research studies on Bitcoin price prediction. This holistic approach should explain whether Bitcoin is a financial asset. Additionally, we also study and compare RNN models with traditional machine learning models and propose a GRU architecture to predict Bitcoin price. GRU's train faster than traditional RNN or LSTM and have not been investigated in the past for cryptocurrency price prediction. In particular, we developed a GRU architecture which can learn the Bitcoin price fluctuations more efficiently than the traditional LSTM. We compare our model to both the traditional neural networks and time series models with different lookback periods to check the robustness of the architecture. For application purposes in algorithmic trading, we have implemented our proposed architecture to test two simple trading strategies for profitability.

## 2. Methodology

A survey of the current literature on neural networks, reveals that traditional neural networks have shortcomings in effectively using prior information for future predictions (Wang, 2015). RNN is a class of neural networks which uses their internal state memory for processing sequences. However, RNNs' on their own are not capable of learning long-term dependencies and they often suffer from short-term memory. With long sequences, especially in time series modelling and textual analysis, RNNs' suffer from vanishing gradient problem during back propagation (Sepp Hochreiter, 1998, Razvan Pascano 2013). If the gradient value shrinks to a very small value, then the RNNs' fail to learn longer past sequences, thus having short-term memory. LSTM, the Long Short-Term Memory (Sepp Hochreiter, 1997), is an RNN architecture with feedback connections, designed to regulate the flow of information. LSTMs' are a variant of the RNN which are explicitly designed to learn long-term dependencies. A single LSTM unit is composed of an input gate, a cell, a forget gate (sigmoid layer and a tanh layer) and an output gate (Figure 1). The gates control the flow of information in and out the LSTM cell. LSTMs' are best suited for time-series forecasting. In the forget gate, the input from the previous hidden state is passed through a sigmoid function along with the input from the current state to generate forget gate output $f_t$. The sigmoid function regulates values between 0 and 1; values closer to 0 are discarded and only values closer to 1 are considered. The input gate is used to update the cell state. Values from the previous hidden state and current state are simultaneously passed through a sigmoid function and a tanh function and the output ($i_t$ and $\tilde{c}_t$) from the two activation functions are multiplied. In this process, the sigmoid function decides which information is important to keep from the tanh output.

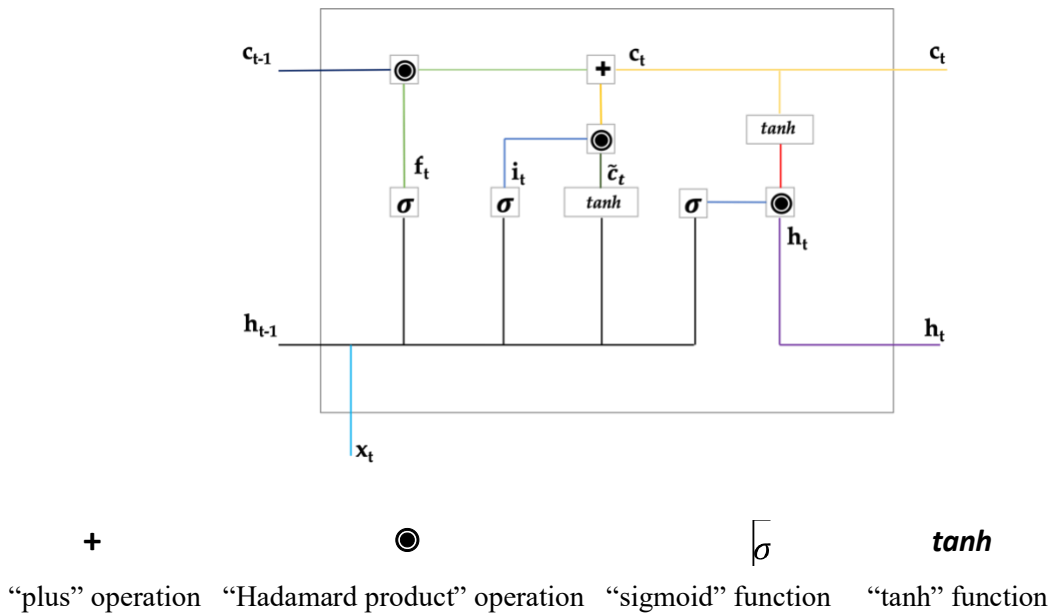

\+      ◉      σ      tanh

"plus" operation   "Hadamard product" operation   "sigmoid" function   "tanh" function

Figure 1. Architecture of a LSTM cell

The previous cell state value is multiplied with the forget gate output and then added pointwise with the output from the input gate to generate the new cell state $c_t$ as shown in equation 1. The output gate operation consists of two steps: first the previous hidden state and current input values are passed through a sigmoid function and secondly the last obtained cell state values are passed through a tanh function. Finally, the tanh output and the sigmoid output are multiplied to produce the new hidden state, which is carried over to the next step. Thus, the forget gate, input gate and output gate decide what information to forget, what information to add from the current step and what information to carry forward respectively. GRU introduced by Cho et. al (2014) solves the problem of the vanishing gradient with a standard RNN. The GRU is similar to LSTM but it combines the forget and the input gates of the LSTM into a single update gate. The GRU further merges the cell state and the hidden state. A GRU unit consists of a cell containing multiple operations which are repeated and each of the operations could be a neural network. Figure 2 below shows the structure of a GRU unit consisting of an update gate, reset gate and a current memory content. These gates enable a GRU unit to store values in the memory for a certain amount of time and use these values to carry information forward, when required, to the current state to update at a future date. In figure 2 below, the update gate is represented by $z_t$ where at each step the input $x_t$ and the output from the previous unit $h_{t-1}$ are multiplied by the weight $W_z$ and added together, and a sigmoid function is applied to get an output between 0 and 1. The update gate addresses the vanishing gradient problem as the model learns how much information to pass forward. The reset gate is represented by $r_t$ in equation 2, where a similar operation as input gate is carried out but this gate in the model is used to determine how much of the past information to forget. The current memory content is denoted by $h'_t$ where $x_t$ is multiplied by $W$ and $r_t$ is multiplied by $h_{t-1}$ element wise (Hadamard product operation) to pass only the relevant information. Finally, a tanh activation function is applied to the summation. The final memory in the GRU unit is denoted by $h_t$ which holds the information for the current unit and passes it on to the network. The computation in the final step is given in equation 2 below. As shown in equation 2, if $z_t$ is close to 0 ((1- $z_t$) close to 1), then most of the current content will be irrelevant and the network will pass majority of the past information and vice versa.

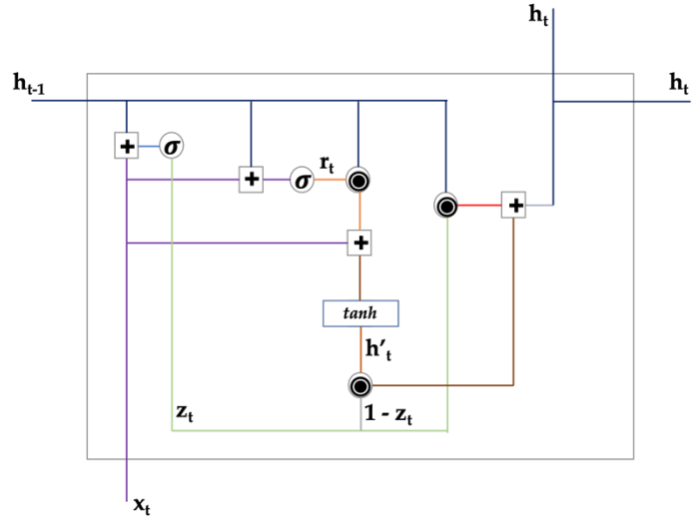

Figure 2. Architecture of a GRU unit

$$z_t = \sigma(W_z.[h_{t-1}, x_t])$$
$$r_t = \sigma(W_r.[h_{t-1}, x_t])$$
$$\tilde{h}_t = \tanh(W.[r_t.h_{t-1}, x_t])$$
$$h_t = (1 - z_t).h_{t-1} + z_t.h_t \tag{2}$$

Both LSTM and GRU are efficient at addressing the problem of vanishing gradient that occurs in long sequence models. GRU's have fewer tensor operations and are speedier to train than LSTMs' (Junyoung Chung, 2014). The neural network models considered for the Bitcoin price prediction are simple neural network (NN), LSTM and GRU. The neural networks are trained with optimized hyperparameters and tested on the test set. A seasonal autoregressive integrated moving average (SARIMA) is considered for comparison purposes. Finally, the best performing model is considered for portfolio strategy execution.

## 3. Data collection and Feature Engineering

Data for the present study has been collected from several sources. We have selected features that may be driving Bitcoin prices and have performed feature engineering to obtain independent variables for future price prediction. Bitcoin prices are driven by a combination of various endogenous and exogenous factors

(Elie Bouri, 2017). The inherent technical aspects of Bitcoin are considered to be endogenous factors. Bitcoin time series data in USD is obtained from bitcoincharts.com. The key features considered in the present study are Bitcoin price, Bitcoin daily lag returns, price volatility, miners revenue, transaction volume, transaction fees, hash rate, money supply, block size and Metcalfe-UTXO. The exogenous features consist of the broader economic and financial indicators that may impact the prices. The exogenous factors of interest in this study are interest rates in the US treasury bond-yields, gold price, VIX volatility index, S&P dollar returns. US treasury bonds and VIX volatility data are used is to investigate the characteristics of Bitcoin investors. Moving average convergence divergence (MACD) are is constructed to explore how moving averages can predict future Bitcoin prices. The roles of Bitcoin as a financial asset, medium of exchange and as a hedge have been studied in the past (Selmi, 2018; Dyhrberg, 2016). Dyhrberg , 2016 prove that there are several similarities of Bitcoin with that of gold and dollar indicating hedging capabilities. Selmi et. al, 2018 study the role of Bitcoin and gold in hedging against oil price movements and conclude that Bitcoin can be used for diversification and for risk management purposes.

The speculative bubble in cryptocurrency markets are often driven by internet search and regulatory actions by different countries (Primavera De Filippi, 2014). In that aspect, internet search data can be considered important for predicting Bitcoin future prices (Eng-Tuck Cheah, 2015; Aaron Yelowitz, 2015). Search data is obtained from Google trends for the keyword "bitcoin". Price of cryptocurrency Ripple (XRP) which is the third biggest cryptocurrency in terms of market cap is also considered as an exogenous factor for Bitcoin price prediction (Efe Caglar Cagli, 2019). Bitcoin data for all the exogenous and endogenous factors for the period 01/01/2010 to 06/30/2019 is collected and a total of 3469 time series observations are obtained. We provide the definitions of these 20 features and the data sources in **Appendix I.**

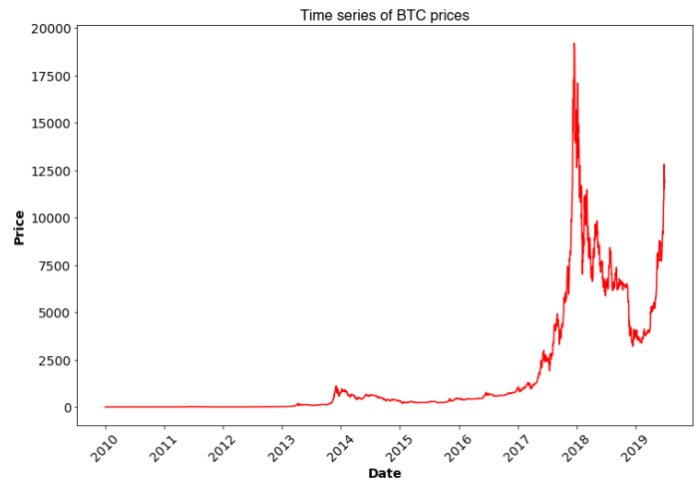

Figure 3. Time series plot of BTC price.

*3.1. Data pre-processing*

Data from different sources are merged and certain assumptions are made. Since, cryptocurrencies get traded twenty-four hours a day and seven days a week, we set the end of day price at 12 AM midnight for each trading day. It is also assumed that the stock, bond and commodity prices maintain Friday's price on the weekends thus ignoring after-market trading. The data-set values are normalized by first demeaning each data-series and then dividing it by its standard deviation. After normalizing the data, the dataset is divided into a training set: observations between Jan 1, 2010 - June 30, 2018, a validation set: observations between July 1, 2018 - Dec 31, 2018 and a test set: observations between Jan 1, 2019 - June 30, 2019. A lookback period of 15, 30, 45 and 60 days are considered to predict the future one-day price and the returns are evaluated accordingly.

*3.2. Feature Selection*

One of the most important aspects of data mining process is feature selection. Feature selection is basically concerned with extracting useful features/patterns from data to make it easier for machine learning models to perform their predictions. To check the behaviour of the features with respect to Bitcoin prices, we plot the data for all the 20 features for the entire time period as shown in figure 4 below. A closer look at the plot reveals that the endogenous features are more correlated with Bitcoin prices than the exogenous features. For the exogenous features, Google trends, interest-rates, and Ripple price seems to be the most correlated.

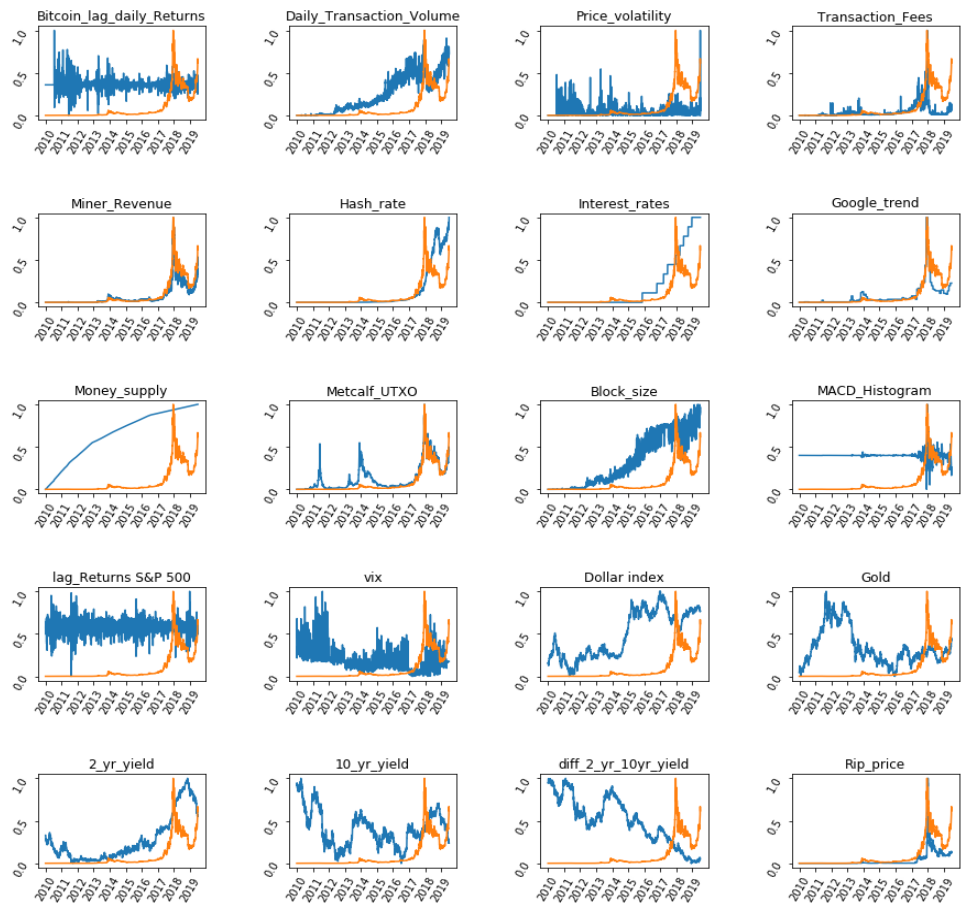

Figure 4. Plot showing the behaviour of independent variables with Bitcoin price. The blue line plots the different features used for Bitcoin price prediction and the orange line plots the Bitcoin price over time.

Multicollinearity is often an issue in statistical learning when the features are highly correlated among themselves and thus the final prediction output is based on very less number of features, which may lead to biased inferences (Kazumitsu Nawata, 2007). To find the most appropriate features for Bitcoin price prediction, a correlation analysis between Bitcoin price and rest of the features is performed. Spearman rank correlation (Spearman, 1904) computed for all the features with Bitcoin price as shown in figure 5 below. We exclude features with a correlation coefficient greater than or equal to 0.8 in our analysis. A set of 15 features are finally selected after dropping Bitcoin Miner Revenue, Metcalf-UTXO, Interest rates, Block size and US Bond yields 2-years and 10-years difference.

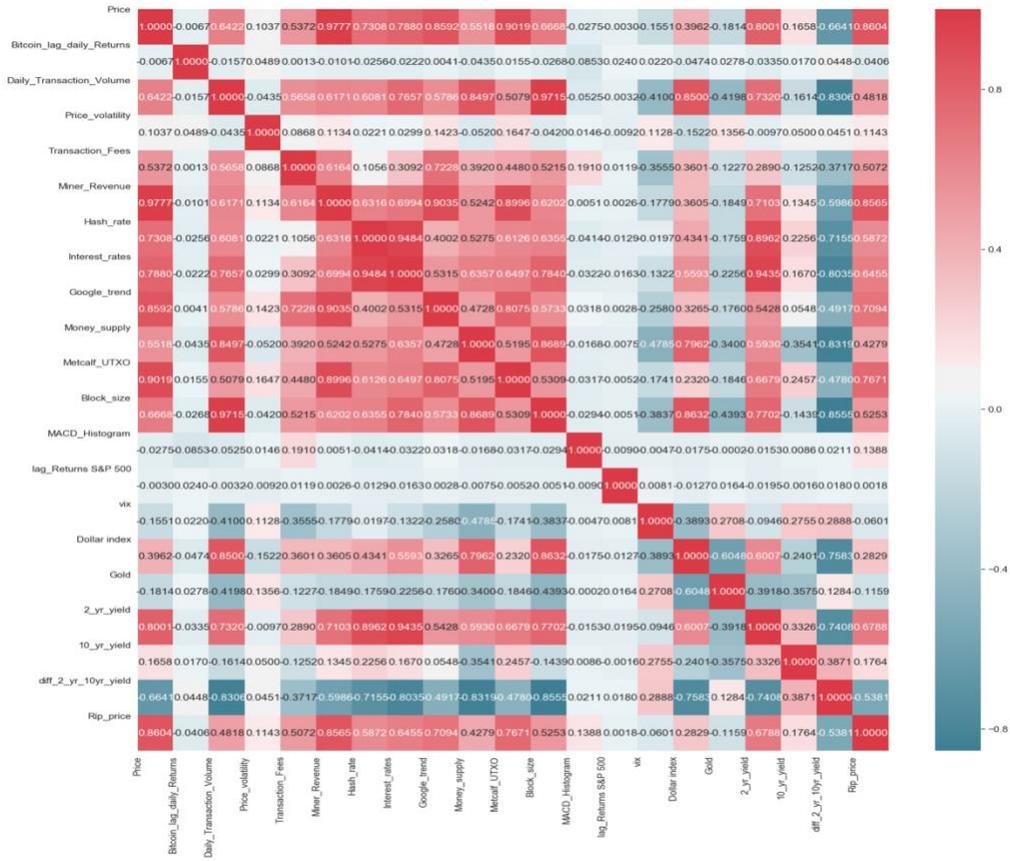

Figure 5. Correlation among the different feature variables used for BTC price prediction.

## 4. Model Implementation and Results

Bitcoin prices follow a time series sequence and so forecasting with standard time series models and comparison with machine learning models are considered. This approach serves two purposes: firstly, time series forecasting will be evaluated based on regression modelling and secondly the relative prediction power of the shallow/deep learning models versus traditional models can be judged. The Bitcoin price graph in figure 3 appears to be non-stationary with an element of seasonality and trend, hence a SARIMA model is considered as one of the baseline models (Paulo Cortez, 2004). At first a simple NN architecture is trained to explore the prediction power of non-linear architectures. A set of shallow learning models is then used to predict the Bitcoin prices using various variants of the RNN as described in section 2. RNN with a LSTM and GRU with dropout and recurrent dropouts are trained and implemented. Keras package (Francois Chollet, 2015) is used with Python 3.6 to build, train and analyse the models on the test set. Deep learning model implementation approach to forecasting is a trade-off between bias and variance the two main source of forecast errors (Yu 2006). Bias error is attributed to inappropriate data assumptions while variance error is attributed to model data sensitivity(Yu 2006). A low variance high bias model leads to underfitting while a low bias high variance model leads to overfitting (Lawrence, 1997). Hence, the forecasting approach aims to find an optimum balance between bias and variance to simultaneously achieve low bias and low variance. In the present study, a high training loss denotes a higher bias while a higher validation loss represents a higher variance. Root mean squared error (RMSE) is preferred over mean absolute error (MAE) for model error evaluation because RMSE gives relatively high weight to large errors.

Each of the individual models are optimized with hyperparameter tuning for price prediction. The main hyperparameters which require subjective inputs are the learning rate alpha, number of iterations, number of hidden layers, choice of activation function, number of input nodes, drop-out ratio and batch-size. A set of activation functions are tested and hyperbolic tangent (TanH) is chosen for optimal learning based on the RMSE error on the test set. TanH suffers from vanishing gradient problem, however the second derivative can sustain for a long time before converging to zero unlike the rectified linear unit (ReLU) which improves RNN model prediction. Initially, the temporal length i.e. the look-back period is taken to be 30 days for the RNN models. The 30-day period is kept in consonance with the standard trading calendar of a month for investment portfolios. Additionally, the best models are also evaluated with a lookback period for 15, 45 and 60 days. Learning rate is one of the most important hyperparameters that can effectively be used for bias-variance trade-off. However, not much improvement in training is observed by altering the learning rate, thus the default value in the Keras package (Francois Chollet, 2015) is used. We train all the models with adam optimization method (Kingma D, Ba Ji, 2015).

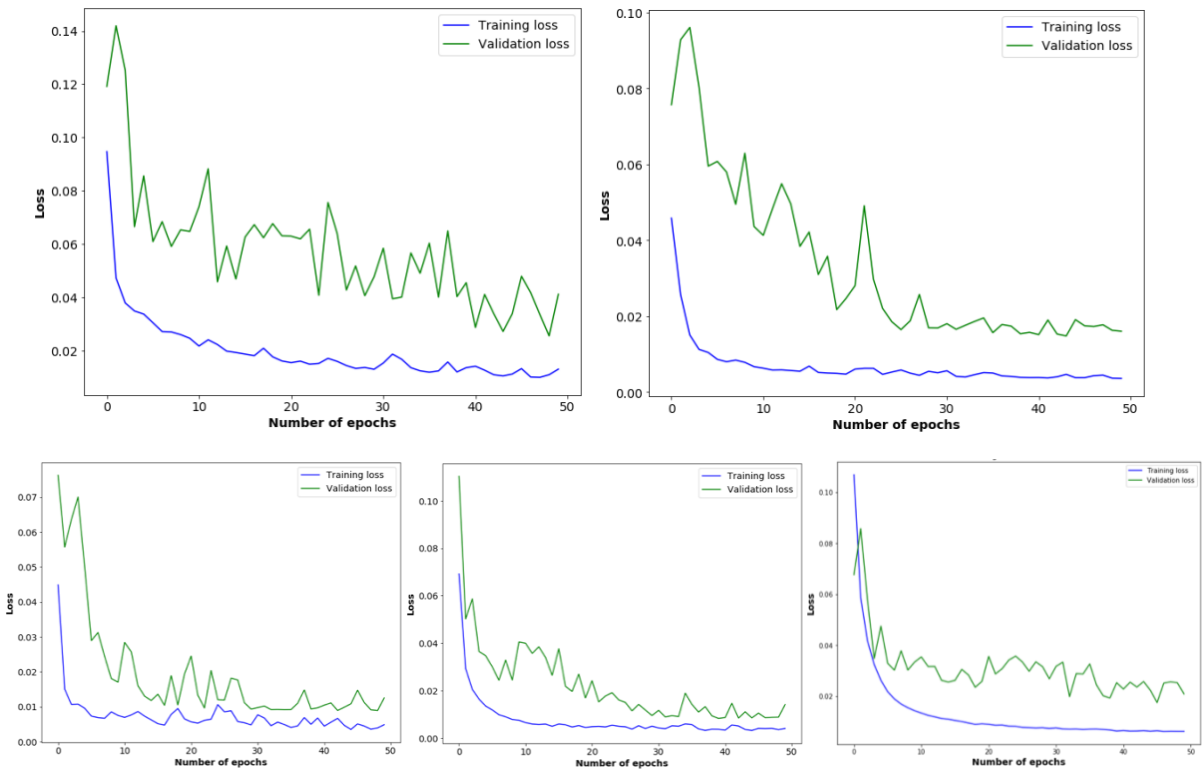

Figure 6. Training and validation loss for simple NN (top left), LSTM with dropout (top right), GRU (bottom left), GRU with a recurrent dropout (bottom middle) and GRU with dropout and recurrent dropout (bottom right).

To reduce complex co-adaptations in the hidden units resulting in overfitting (Srivastava et. al, 2014), dropout is introduced in the LSTM and GRU layers. Thus, for each training sample the network is re-adjusted and a new set of neurons are dropped out. For both LSTM and GRU architecture, a recurrent dropout rate (Gal, Y. & Ghahramani, Z., 2016) of 0.1 is used. For the two hidden layers GRU, a dropout of 0.1 is additionally used along with the recurrent dropout of 0.1. The dropout and recurrent dropout rates are optimized to ensure that the training data is large enough to not to be memorized in spite of the noise and to avoid overfitting (Srivastava et. al, 2014). For the simple NN, two dense layers are used each with hidden nodes 25 and 1. The LSTM layer is modelled with one LSTM layer (50 nodes) and one dense layer (1 node). The simple GRU and the GRU with recurrent dropout architecture comprised of one GRU layer (50 nodes) and one dense layer with 1 node. The final GRU architecture is tuned with two GRU layers (50 nodes and 10 nodes) with a dropout and recurrent dropout of 0.1. The optimized batch size for the neural network and the RNN models are determined to be 125 and 100 respectively. A higher batch size led to a higher training and validation loss during the learning process.

Figure 6 shows the training and validation loss for the neural network models. The difference between training loss and validation loss reduces with a dropout and a recurrent dropout for the one GRU layer model (Figure 6 bottom middle). However, with addition of an extra GRU layer, the difference between the training and validation loss increases. After training, all the neural network models are tested on the test data along with the SARIMA model. The RMSE for all the models on the train and test data are shown in Table 1. As seen from Table 1, the LSTM architecture performs better than the simple NN architecture due to memory retention capabilities (Sepp Hochreiter, 1997). As seen from Table I, the GRU model with a recurrent dropout generates an RMSE of 0.014 on the training set and 0.017 on the test set. RNN-GRU performs better than LSTM and a plausible explanation is the fact that GRUs are computationally faster with lesser number of gates and tensor operations. The GRU controls the flow of information like the LSTM unit, however the GRU has no memory unit and it exposes the full hidden content without any control (Chung, 2014). GRUs also tend to perform better than LSTM on less training data (Łukasz Kaiser & Ilya Sutskever, 2016) as in the present case while LSTMs are more efficient in remembering longer sequences (Wenpeng Yin, 2017). We also find that the recurrent dropout in the GRU layer helps reduce the RMSE on the test data and the difference of RMSE between training and test data is the minimum for the GRU model with recurrent dropout. These results indicate that the GRU with recurrent dropout is the best performing model for our problem. Recurrent dropouts help to mask some of the output from the first GRU layer which can be thought as a variational inference in RNN (Gal 2016, Merity 2017). We have also trained the GRU recurrent dropout model with a lookback period of 15, 45 and 60 days and the results are reported in Table 2. It can be concluded from Table 2 that the lookback period for 30 days is the optimal period for the best RMSE results. Figure 7 shows the GRU model with recurrent dropout predicted Bitcoin price in the test data as compared to the original data. The model predicted price is higher than the original price in the first few months of 2019, however when Bitcoin price shot up in June – July 2019, the model is able to learn this trend effectively.

Table 1: Train test RMSE of 30 days lookback period for different models

| Models | RMSE Train | RMSE Test |
|---|---|---|
| Neural Network | 0.020 | 0.031 |
| LSTM | 0.010 | 0.024 |
| GRU | 0.010 | 0.019 |
| GRU-Dropout | 0.014 | 0.017 |
| GRU-Dropout-GRU | 0.012 | 0.034 |
| SARIMA | 0.034 | 0.041 |

Table 2: Train test RMSE for GRU recurrent model

| Lookback Period (days) | RMSE Train | RMSE Test |
|---|---|---|
| 15 | 0.012 | 0.016 |
| 45 | 0.011 | 0.019 |
| 60 | 0.011 | 0.017 |

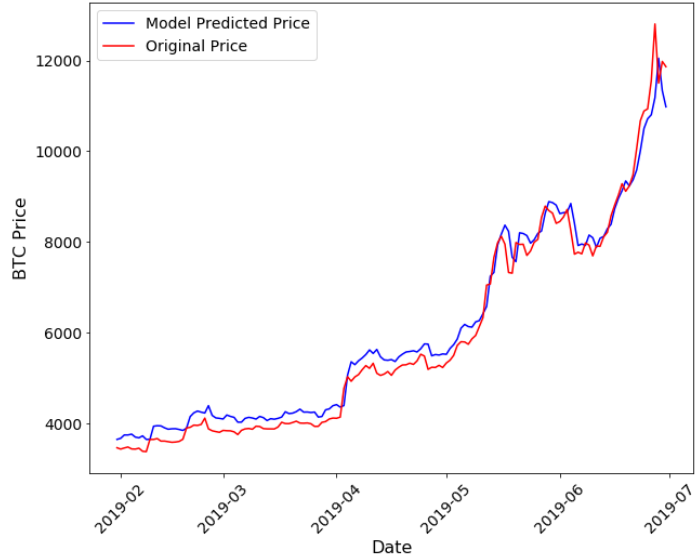

Figure 7. Bitcoin price as predicted by the GRU one-layer model with dropout and recurrent dropout.

**5. Portfolio Strategy**

We implement two trading strategies to evaluate our results in portfolio management of cryptocurrencies. For simplicity we consider only Bitcoin trading and we assume that the trader only buys and sells based on the signals derived from quantitative models. Based on our test set evaluation, we have considered the GRU one layer with recurrent dropout as our best model for implementing the trading strategies. Two types of trading strategies are implemented as discussed in this section. The first strategy is a long-short strategy wherein the buy signal predicted from the model will lead to buying the Bitcoin and a sell signal will essentially lead to short-selling the Bitcoin at the beginning of the day based on the model predictions for that day. If the model predicted price on a given day is lower than the previous day, then the trader will short sell the Bitcoin and cover them at the end of the day. An initial portfolio value of 1 is considered and the transaction fees is taken to be 0.8% of the invested or sold amount. Due to daily settlement, the long-short strategy is expected to incur significant transaction costs which may reduce the portfolio value. The second strategy is a buy-sell strategy where the trader goes long when a buy signal is triggered and sell all the bitcoins when a sell signal is generated. Once the trader sells all the coins in the portfolio, he/she waits for the next positive signal to invest again. When a buy signal occurs, the trader invests in Bitcoin and remains invested till the next sell signal is generated.

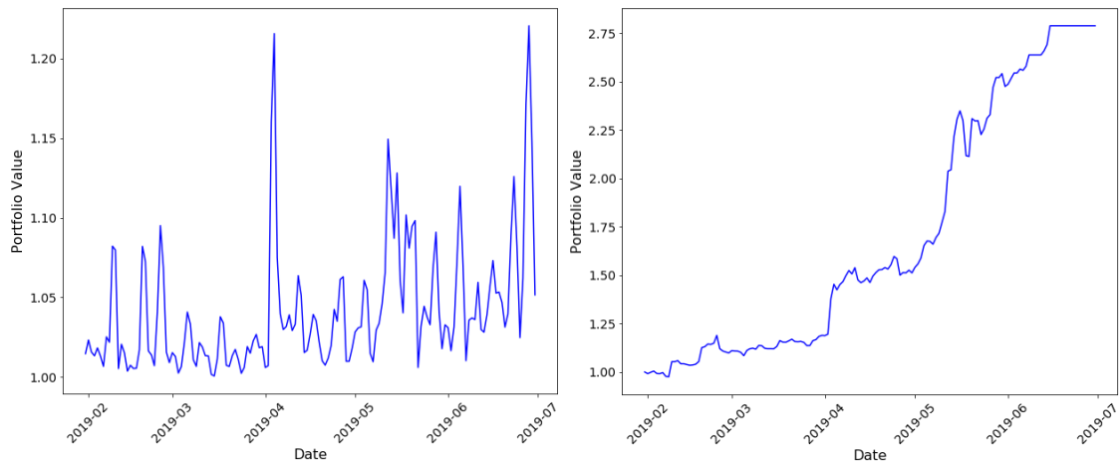

Figure 8 above shows the change in portfolio value over time when the strategies long-short (left) and buy-sell (right) are implemented on the test data. Due to short selling, daily settlement the long-short portfolio incurs transaction fees which reduces growth and increases volatility in the portfolio.

Most Bitcoin exchanges, unlike stock exchanges, do now allow short selling of Bitcoin yet, due to higher volatility and regulatory risks (Primavera De Filippi, 2014). Also volatility depends on how close the model predictions are to the actual market price of Bitcoin at every point of time. As can be seen from Figure 8, Bitcoin prices go down during early June 2019, and the buy-sell strategy correctly predicts the fall, with the trader selling the Bitcoins holding to keep the cash before investing again when the price starts rising from mid-June. In comparison, due to short selling and taking long positions simultaneously, the long-short strategy suffers during the same period of time with very slow increase in portfolio value. However, long-short strategies might be more powerful when we consider a portfolio consisting of multiple cryptocurrencies where investors can take simultaneous long and short positions in currencies which have significant growth potential and overvalued currencies.

## 6. Conclusions

There have been considerable number of studies on Bitcoin price prediction using machine learning and time-series analysis (Wang 2015). However, most of these studies have been mostly based on predicting the Bitcoin prices based on pre-decided models with limited number of features. The present study explores Bitcoin price prediction based on a collective and exhaustive list of features with financial linkages. The basis of any investment has always been wealth creation either through fundamental investment or technical speculation, and cryptocurrencies are no exception to this. In this study, feature engineering is performed taking into account whether Bitcoin could be used as an alternative investment that offers investors diversification benefits and a different investment avenue when the traditional means of investment are not doing well. This study considers a holistic approach to select the predictor variables that might be helpful in learning future Bitcoin price trends. The US treasury 2-year and 10-year yields are the benchmark indicators for short-term and long-term investment in bond markets, hence a change in these benchmarks could very well propel investors towards alternative investment avenues such as the Bitcoin. Similar methodology can be undertaken for Gold, S&P returns and Dollar Index. Whether it is good news or bad news, increasing attraction or momentum based speculation, google trends, volatility index (VIX), MACD, lagged prices data are perfect for studying this aspect of the influence on the prices.

We also conclude that recurrent neural network models such as LSTM and GRU outperform conventional time-series approaches like SARIMA for price prediction. With limited data, neural networks like LSTM and GRU can regulate past information to learn effectively from non-linear patterns. Deep models require accurate training and hyperparameter tuning to yield results which might be computationally extensive for large datasets unlike conventional time-series approaches. However, for stock price prediction or cryptocurrency price prediction, the market data is always limited and computational complexity is not a concern and thus shallow learning models can be effectively used in practice. These benefits will likely contribute significantly to quantitative finance in the coming years.

In deep learning literature, LSTM has been traditionally used to analyse time-series. GRU architecture on the other hand, seems to be performing better than the LSTM model in our analysis. The simplicity of the GRU model, where the forgetting and updating is occurring simultaneously, is found to be working well in Bitcoin price prediction. Adding a recurrent dropout improves the performance of the GRU architecture, however further studies need to be undertaken to explore the dropout phenomenon in GRU architectures. Two types of investment strategies have been implemented with our trained GRU architecture. Results show that when machine learning models are implemented with full understanding, it can be beneficial to the investment industry for financial gains and portfolio management. In the present case, machine learning models perform much better than traditional SARIMA in price prediction; thus making the investment strategies valuable. With proper back testing of each of these models can contribute to manage portfolio risk and reduce financial losses. Nonetheless a significant improvement over the current study can be achieved if a bigger data set is available.

**Author Contributions:** All authors have read and agree to the published version of the manuscript. Conceptualization, data curation, validation and draft writing, S.K.; Methodology, formal analysis, draft preparation and editing, A.D.; coding, plots, M.B.

**Funding:** This research received no external funding.

**Acknowledgments:** The authors would like to thank the staff at Haas School of Business, University of California Berkeley and Katz School of Business, University of Pittsburgh for their support.

**Conflicts of Interest:** The authors declare no conflict of interest.

## Appendix A

**Definition of variables and data source**

| Features | Definition | Source |
|---|---|---|
| Bitcoin Price | Bitcoin prices | https://charts.bitcoin.com/btc/ |
| BTC Price Volatility | The annualized daily volatility of price changes. Price volatility is computed as the standard deviation of daily returns, scaled by the square root of 365 to annualize, and expressed as a decimal. | https://charts.bitcoin.com/btc/ |
| BTC Miner Revenue | Total value of Coinbase block rewards and transaction fees paid to miners. Historical data showing (number of bitcoins mined per day + transaction fees) * market price. | https://www.quandl.com/data/BCHAIN/MIREV-Bitcoin-Miners-Revenue |
| BTC Transaction volume | The number of transactions included in the blockchain each day | https://charts.bitcoin.com/btc/ |
| Transaction Fees | Total amount of Bitcoin Core (BTC) fees earned by all miners in 24-hour period, measured in Bitcoin Core (BTC). | https://charts.bitcoin.com/btc/ |

| Hash Rate | The number of block solutions computed per second by all miners on the network | https://charts.bitcoin.com/btc/ |
|---|---|---|
| Money Supply | The amount of Bitcoin Core (BTC) in circulation | https://charts.bitcoin.com/btc/ |
| Metcalfe-UTXO | Metcalfe's Law states that the value of a network is proportional to the square of the number of participants in the network. | https://charts.bitcoin.com/btc/ |
| Block size | Miners collect Bitcoin Core (BTC) transactions into distinct packets of data called blocks. Each block is cryptographically linked to the preceding block, forming a "blockchain." As more people use the Bitcoin Core (BTC) network for Bitcoin Core (BTC) transactions, the block size increases | https://charts.bitcoin.com/btc/ |
| Google Trends | This is the month-wise google search results for the Bitcoins. | https://trends.google.com |
| Volatility (VIX) | VIX, is a real-time market index that represents the market's expectation of 30-day forward-looking volatility | http://www.cboe.com/products/vix-index-volatility/vix-options-and-futures/vix-index/vix-historical-data |
| Gold price level | Gold Price level | https://www.quandl.com/data/WGC/GOLD_DAILY_USD-Gold-Prices-Daily-Currency-USD |
| US Dollar Index | The U.S. dollar index (USDX) is a measure of the value of the U.S. dollar relative to the value of a basket of currencies of the majority of the U.S.'s most significant trading partners. | https://finance.yahoo.com/quote/DX-Y.NYB/history?period1=1262332800&period2=1561878000&interval=1d&filter=history&frequency=1d |
| US Bond yields | 2-year / short-term yields | https://www.quandl.com/data/USTREASURY/YIELD-Treasury-Yield-Curve-Rates |
| US Bond yields | 10-year/ long term yields | https://www.quandl.com/data/USTREASURY/YIELD-Treasury-Yield-Curve-Rates |
| US Bond yields | Difference between 2 year and 10 year / synonymous with yield inversion and recession prediction | https://www.quandl.com/data/USTREASURY/YIELD-Treasury-Yield-Curve-Rates |
| MACD | MACD=12-Period EMA −26-Period EMA. We have taken the data of the MACD with the signal line.<br>MACD line = 12 day EMA Minus 26 day EMA<br>Signal line = 9 day EMA of MACD line<br>MACD Histogram = MACD line Minus Signal line | |
| Ripple Price | The price of an alternative cryptocurrency | https://coinmarketcap.com/currencies/ripple/historical-data/?start=20130428&end=20190924 |
| One day Lagged S&P 500 Market returns | Stock market returns | https://finance.yahoo.com/quote/%5EGSPC/history?period1=1230796800&period2=1568012400&interval=1d&filter=history&frequency=1d |
| Interest rates | The federal funds rate decide the shape of the future interest rates in the economy. | http://www.fedprimerate.com/fedfundsrate/federal_funds_rate_history.htm#current |